\documentclass[journal]{IEEEtran}

\ifCLASSINFOpdf
\else
   \usepackage[dvips]{graphicx}
\fi
\usepackage{url}
\usepackage{graphicx}
\usepackage{booktabs}
\usepackage{microtype}
\usepackage{amsmath}
\providecommand{\singlepagebreak}{\clearpage}

\begin{document}

\title{Low-Latency Neural Models for \\ Real-Time Music Enhancement}

\author{ { Emmanouil Karystinaios \quad Jonathan Greif \quad David Nadrchal \quad Paul Primus\quad Gerhard Widmer }\\
\IEEEmembership{Institute of Computational Perception, Johannes Kepler University Linz, Austria} \\ 
\{firstname.lastname\}@jku.at
}

\maketitle

\begin{abstract}
Music recordings and live streams are often affected by noise, reverberation, spectral imbalances, or artifacts that degrade listening quality. While speech enhancement has matured into a well-defined research area, music enhancement is less established because musical signals combine overlapping sources, wide bandwidths, strong dynamics, and intentional production effects. We study real-time music enhancement under strict causal and low-latency constraints. We formulate the task around recovery of the intended produced mix from acoustic and production-oriented degradations, adapt compact causal networks to music, and compare speech-derived real-time baselines, an external music-denoising model, an offline restoration reference, and a music-specific MusicFilterNet-MS variant. On the tested hardware, all causal models run faster than real time, but improvements depend strongly on the dataset, degradation type, and metric family; under several objective criteria, indiscriminate enhancement can worsen the degraded input. The main contribution is therefore a benchmark and an analysis rather than a universal best model: real-time music enhancement is feasible, but robust improvement requires degradation-aware modeling, stereo-aware processing, identity-preserving correction, and evaluation beyond a single objective score.
\end{abstract}

\section{Introduction}
\label{sec:intro}

Enhancement of degraded audio has been studied for decades, most prominently in the context of speech. Speech enhancement systems target additive noise and reverberation, with well-established datasets and metrics supporting progress in telecommunications, hearing aids, and interactive devices~\cite{drgas2023survey}. In contrast, music enhancement is less clearly defined. Music signals combine multiple overlapping sources, broader spectral ranges, and intentional nonlinearities from production, making them more complex to model and evaluate.

Nevertheless, there are clear application scenarios where music enhancement is valuable. These include at-home recording and online streaming under non-ideal acoustic conditions, as well as on-device live capture of performances (with or without video) and real-time rehearsals in suboptimal conditions. In such cases, listeners may expect cleaner and more balanced signals, with reduced noise, room reverberation, or artifacts, and minimal latency to enable real-time use.

Related work in music processing has addressed specific aspects such as denoising and restoration of historical recordings~\cite{moliner2022two,moliner2022behm} and dedicated denoising networks for modern recordings~\cite{cheng2024musicecan}. More recently, SonicMaster~\cite{melechovsky2025sonicmaster} proposed an all-in-one approach to music restoration and mastering, introducing a taxonomy of degradations, including equalization, dynamics, reverberation, amplitude defects, and stereo-image issues, with dedicated evaluation metrics, but only in offline settings.

Real-time approaches to music denoising, restoration, and enhancement are limited. By contrast, real-time speech enhancement has advanced considerably, with causal and low-latency neural architectures now widely deployed~\cite{crn,hu2020dccrn,schroter2023deepfilternet}. Extending these techniques to music is nontrivial because of higher source density, stronger dynamics, and perceptual criteria that differ from intelligibility-driven speech tasks.

In this paper, we take a step toward real-time music enhancement under a shared 44.1~kHz causal streaming protocol. Our novelty lies in jointly adapting the training objective, output parameterization, and model design to music. Specifically, we contribute: (i) a degradation formulation that distinguishes the intended produced mix from subsequent acoustic and production-oriented corruption; (ii) a three-stage curriculum that progresses from multi-resolution spectral reconstruction to adversarial refinement and then to a new music-oriented composite objective with waveform, log-mel, instantaneous-frequency, and level-preservation terms; (iii) music-specific adaptations of compact causal networks, including mel-domain processing and identity-centered residual correction, together with the new MusicFilterNet-MS architecture; and (iv) a benchmark using quality metrics and operational measurements.\footnote{All code is available at: \url{https://github.com/manoskary/audio-enhancement}}

We emphasize that our goal is not to match large offline systems such as SonicMaster, which are non-causal and use long contexts. Instead, we use SonicMaster as an offline reference and focus on three questions under strict real-time constraints: (i) which compact causal architectures remain computationally viable, (ii) how far speech-derived real-time models transfer to music, and (iii) which degradation categories and metric families expose the remaining failure modes. The results do not support a universal state-of-the-art claim or identify MFN-MS as the overall best model. Rather, they show that causal music enhancement is computationally feasible, but improvement is degradation- and metric-dependent, and an always-on global correction can be less faithful than leaving the input unchanged. This motivates degradation-aware routing, stereo-aware processing, and an identity-preserving fallback when correction is uncertain.

\begin{figure*}[t]
\centering
\includegraphics[width=0.8\linewidth]{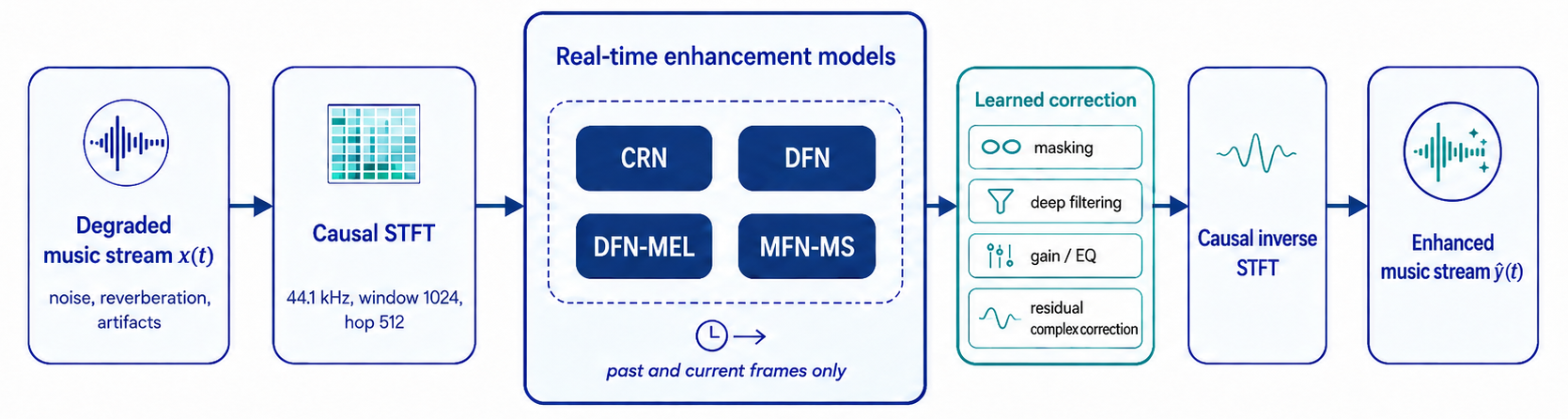}
\caption{Framework of the real-time music enhancement pipeline. All evaluated neural models operate on causal STFT frames and are benchmarked at a batch size of one with the same analysis/synthesis latency.}
\label{fig:framework}
\end{figure*}

\section{Problem Formulation and Novelty}

\subsection{Signal Model for Music Enhancement}

Let $\mathbf{y}$ denote the intended $C$-channel produced mix, where $C=1$ for mono and $C=2$ for stereo. We write
\begin{equation}
\mathbf{y} = \mathcal{P}(s_1,\ldots,s_M),
\qquad
\mathbf{x} = \mathcal{D}_{\phi}(\mathbf{y},\mathbf{n}),
\label{eq:general_degradation}
\end{equation}
where $\mathcal{P}$ denotes the artistic mixing and production process, and $\mathcal{D}_{\phi}$ denotes a possibly nonlinear, time-varying, and channel-dependent degradation process. The latter may include room or device filtering, additive interference $\mathbf{n}$, additional equalization or dynamics processing, clipping, amplitude defects, quantization or codec distortion, and stereo-image transformations. An operation belongs to $\mathcal{P}$ when it is present in the reference production, and to $\mathcal{D}_{\phi}$ when it is subsequently applied to construct the degraded observation.

For the linear, time-invariant, additive special case,
$\mathcal{D}_{\phi}(\mathbf{y},\mathbf{n})
= \mathbf{h} * \mathbf{y} + \mathbf{n}$, and the corresponding
STFT-domain relation is
\begin{equation}
\mathbf{X}(k,f)
\approx
\mathbf{H}(f)\mathbf{Y}(k,f) + \mathbf{N}(k,f),
\label{eq:lti_degradation}
\end{equation}
up to STFT windowing effects. Here, $\mathbf{H}(f)$ may represent either a scalar transfer function or a multichannel transfer matrix. Nonlinear and channel-coupled degradations remain represented by $\mathcal{D}_{\phi}$ and not by an additive artifact term.

A frame-causal enhancement model estimates
$\widehat{\mathbf{Y}}(k,\cdot)
=F_{\theta}\!\left(\{\mathbf{X}(\tau,\cdot)\}_{\tau\leq k}\right)$,
so output frame $k$ depends only on the current and preceding input frames. The target is the produced mix $\mathbf{Y}$ rather than the individual stems, and the objective is to reduce unwanted degradations while preserving intended production characteristics and interchannel relationships. Channel indices are omitted below for readability.

\subsection{FINALLY-Inspired Training Objective}
\label{subsec:loss}

We use a three-stage curriculum inspired by the training progression of FINALLY~\cite{babaev2024finally}, but with different objectives and stage transitions. FINALLY first trains a 16-kHz model with a reconstruction objective, then adds adversarial and feature-matching losses, and finally introduces a 48-kHz upsampling module together with human-feedback losses. In contrast, each of our causal architectures remains fixed at 44.1~kHz across the three stages. We use no speech self-supervised encoder or speech-quality predictor; the final stage instead introduces signal-domain constraints intended for music.

\subsubsection{Stage 1: Spectral reconstruction}
Let $\mathcal{R}$ denote the set of STFT resolutions. We define the multi-resolution spectral reconstruction objective as
\begin{equation}
\mathcal{L}_{MR}
=
\frac{1}{|\mathcal{R}|}
\sum_{r\in\mathcal{R}}
\left(
\lambda_{\mathrm{SC}}\mathrm{SC}_{r}
+
\lambda_{\mathrm{LM}}\mathcal{L}_{\mathrm{LM}}^{(r)}
\right),
\label{eq:mrstft}
\end{equation}
where $\mathrm{SC}_{r}$ is spectral convergence and $\mathcal{L}_{\mathrm{LM}}^{(r)}$ is a log-magnitude distance at resolution $r$.

\subsubsection{Stage 2: Adversarial refinement}
Stage~2 retains the reconstruction objective and introduces a multi-resolution STFT discriminator with hinge adversarial and feature-matching losses:
\begin{equation}
\mathcal{L}_{G}^{(2)}
=
\mathcal{L}_{\mathrm{MR}}
+
\lambda_{\mathrm{adv}}\mathcal{L}_{\mathrm{adv}}^{\mathrm{hinge}}
+
\lambda_{\mathrm{FM}}\mathcal{L}_{\mathrm{FM}}.
\label{eq:stage2}
\end{equation}
Feature matching regularizes the generator through intermediate discriminator activations, while the adversarial term encourages outputs whose time-frequency statistics are consistent with the reference music distribution.

\subsubsection{Stage 3: Music-oriented multi-domain fine-tuning}
Stage~3 replaces the reconstruction branch of
Eq.~\eqref{eq:stage2} with a composite music objective while
retaining adversarial and feature-matching supervision:
\begin{align}
\mathcal{L}_{\mathrm{music}}
&=
\lambda_{\mathrm{si}}\mathcal{L}_{\mathrm{SI}}
+
\lambda_{\mathrm{pe}}\mathcal{L}_{\mathrm{PE\text{-}SI}}
+
\lambda_{\mathrm{l1}}\mathcal{L}_{\mathrm{L1\text{-}dB}}
\nonumber\\
&\quad+
\lambda_{\mathrm{MR}}\mathcal{L}_{\mathrm{MR}}
+
\lambda_{\mathrm{mel}}\mathcal{L}_{\mathrm{logmel}}
+
\lambda_{\mathrm{IF}}\mathcal{L}_{\mathrm{IF}}
+
\lambda_{\mathrm{gain}}\mathcal{L}_{\mathrm{gain}},
\label{eq:music_loss}\\
\mathcal{L}_{G}^{(3)}
&=
\mathcal{L}_{\mathrm{music}}
+
\lambda_{\mathrm{adv}}\mathcal{L}_{\mathrm{adv}}^{\mathrm{hinge}}
+
\lambda_{\mathrm{FM}}\mathcal{L}_{\mathrm{FM}}.
\label{eq:stage3}
\end{align}

Here $\mathcal{L}_{\mathrm{SI}}$ and $\mathcal{L}_{\mathrm{PE\text{-}SI}}$ are scale-invariant waveform losses computed on the original and pre-emphasized signals, respectively. The level-normalized $\mathcal{L}_{\mathrm{L1\text{-}dB}}$ term combines an L1 reconstruction error normalized by the target level with adaptive log-RMS regularization, while $\mathcal{L}_{\mathrm{gain}}$ penalizes framewise log-RMS deviations. The log-mel term constrains the frequency-dependent energy envelope, and $\mathcal{L}_{\mathrm{IF}}$ penalizes differences between wrapped temporal phase increments. The objective therefore combines waveform, spectral, phase-evolution, and level-preservation constraints without using speech-specific perceptual features or quality predictors. Further implementation details and grouped loss ablations are provided in the Supplementary Material.

\begin{table*}[!t]
\centering
\caption{Evaluation on M\&N and SonicMaster. Lower is better for FAD/KL; higher is better for MM-SNR, SI-SNR, SSIM, PQ, and ZIM. Shown are the degraded input, Stage~3 real-time baselines, MFN-MS, and offline references; the complete stage-wise table is in the Supplementary Material. Best non-input values in each dataset section are \textbf{bold}.}
\scriptsize
\setlength{\tabcolsep}{2.6pt}
\renewcommand{\arraystretch}{1.02}
\begin{tabular}{l|rrrrrrr|rrrrrrr}
\toprule
& \multicolumn{7}{c|}{\textbf{M\&N Dataset}} & \multicolumn{7}{c}{\textbf{SonicMaster Dataset}} \\
\cmidrule(lr){2-8}\cmidrule(lr){9-15}
\textbf{Model}
& FAD$\downarrow$ & KL$\downarrow$ & MM$\uparrow$ & SI$\uparrow$ & SSIM$\uparrow$ & PQ$\uparrow$ & ZIM$\uparrow$
& FAD$\downarrow$ & KL$\downarrow$ & MM$\uparrow$ & SI$\uparrow$ & SSIM$\uparrow$ & PQ$\uparrow$ & ZIM$\uparrow$ \\
\midrule
Degraded input
& 0.640 & 0.152 & 5.336 & 3.509 & 0.537 & 5.818 & 2.391
& 0.103 & 0.077 & 8.477 & 0.023 & 0.501 & 7.007 & 3.659 \\
\midrule
CRN Stage~3
& 0.684 & 0.099 & 7.048 & 4.556 & 0.502 & 5.903 & 2.339
& 0.102 & 0.350 & \textbf{5.511} & \textbf{-1.990} & 0.508 & \textbf{6.988} & \textbf{3.657} \\
DFN Stage~3
& 0.655 & 0.184 & 4.001 & 1.929 & 0.465 & 5.742 & 2.338
& 0.114 & 0.205 & 3.296 & -4.410 & 0.503 & 6.869 & 3.528 \\
DFN-MEL Stage~3
& 0.754 & 0.132 & 5.348 & 3.898 & 0.512 & 5.475 & 2.223
& 0.145 & 0.222 & 2.794 & -5.179 & 0.494 & 6.564 & 3.306 \\
MFN-MS Stage~3
& \textbf{0.640} & 0.190 & 4.284 & 2.689 & 0.469 & 5.657 & \textbf{2.428}
& 0.116 & 0.074 & 4.330 & -2.877 & 0.503 & 6.878 & 3.271 \\
MusicECAN (offline)
& 0.644 & \textbf{0.056} & \textbf{8.596} & \textbf{5.532} & \textbf{0.573} & \textbf{5.904} & 2.376
& 0.188 & 0.082 & 4.457 & -3.073 & 0.447 & 6.843 & 3.429 \\
SonicMaster (offline)
& 0.697 & 0.124 & 2.654 & -1.150 & 0.456 & 5.898 & 2.296
& \textbf{0.072} & \textbf{0.008} & 2.021 & -3.309 & \textbf{0.519} & 3.028 & 2.885 \\
\bottomrule
\end{tabular}
\label{tab:mnm_sonicmaster_metrics}
\end{table*}

\subsection{Novel Contributions}
\label{sec:novelty}

Our technical novelty lies in the joint adaptation of the training objective, output parameterization, and causal model design to music enhancement. We introduce: (i) a FINALLY-inspired but music-specific training curriculum; (ii) a composite multi-domain objective with explicit level-preservation components; (iii) mel-domain and identity-centered adaptations of compact causal enhancement networks; and (iv) MusicFilterNet-MS as a music-specific causal architecture. The contribution is therefore broader than a transfer of speech models, while not claiming that every constituent loss is individually new.

\subsubsection{Music-oriented composite objective}
The Stage~3 objective combines established scale-invariant waveform, multi-resolution spectral, log-mel, and instantaneous-frequency constraints with explicit level-aware terms. In particular, the level-normalized L1 term penalizes reconstruction error relative to the target signal level and adds adaptive log-RMS regularization, while the framewise gain-consistency term discourages local loudness drift. The novelty lies in their music-oriented composition and staged integration with adversarial training, rather than in a claim of priority for the standard SI, MR-STFT, log-mel, or phase-difference terms individually.



\subsubsection{Model adaptations}
We evaluate two compact causal baselines from real-time speech enhancement. The first is a convolutional recurrent network (CRN), a causal encoder--recurrent--decoder architecture that predicts time-frequency corrections from past and current frames only. The second is DeepFilterNet (DFN)~\cite{schroter2023deepfilternet}. We use both the original ERB-style configuration and a mel-domain variant (DFN-MEL), in which the original ERB-based representation is replaced by a mel-frequency representation that is more aligned with music timbre.

Following our problem formulation, we do not assume that enhancement can be reduced to predicting purely attenuating spectral masks. In music, intended production effects and degradations may interact in complex ways, requiring both attenuation and amplification. To account for this, we replace the original output heads with an identity-centered residual complex ratio mask
\begin{multline}
  M = (1 + \alpha \tanh r) + j(\alpha \tanh i) \\
\text{applied to the mixture STFT as: } \hat{Y} = M \odot X.
\end{multline}
Here $r$ and $i$ are the real and imaginary outputs of the network head, $\alpha$ bounds each Cartesian component of the residual correction, and $\odot$ denotes element-wise complex multiplication. This parameterization has an exact identity point at $r=i=0$ and permits both attenuation and amplification. Exact passthrough at initialization additionally requires the final output head to be initialized at zero.

\subsubsection{MusicFilterNet-MS}
To test a music-specific causal alternative to direct transfer from speech models, we also introduce MusicFilterNet-MS (MFN-MS). MFN-MS keeps the same streaming STFT interface as the other models but increases music-specific capacity with absolute frequency encodings, a causal local time-frequency encoder, gated deep filtering, smooth equalization and gain heads, a lightweight waveform refiner, and several causal complex residual refinement stages. The design is inspired by the channel-attention and denoising emphasis of MusicECAN~\cite{cheng2024musicecan}, but every temporal operation is causal and the output remains a frame-synchronous real-time enhancement model. Later experiments show that MFN-MS is computationally feasible, but its learned correction is currently only consistently beneficial on some degradation categories; we therefore use it as a music-specific diagnostic model rather than claiming it as the overall best system.

\section{Experimental Setup}
\label{sec:setup}

Our experiments evaluate real-time enhancement models against the degraded input, a music-denoising baseline (MusicECAN~\cite{cheng2024musicecan}), and offline restoration (SonicMaster~\cite{melechovsky2025sonicmaster}). MusicECAN is included to test if a model trained for music denoising transfers beyond additive-noise removal, and SonicMaster is included as a non-causal restoration/mastering reference not as a real-time competitor. 
Prior work~\cite{cheng2024musicecan, moliner2022two} has used a limited set of evaluation measures; we therefore report both generic and music-oriented metrics.

\subsubsection{Perceptual and music-specific metrics}
Following SonicMaster~\cite{melechovsky2025sonicmaster}, we report embedding and spectrogram-based perceptual proxies: Fr{\'{e}}chet Audio Distance (FAD)~\cite{roblek2019fad} computed from CLAP embeddings~\cite{elizalde2023clap}, structural similarity index (SSIM), log-mel KL divergence, and Production Quality (PQ)~\cite{tjandra2025aes}. We also report Zimtohrli~\cite{alakuijala2025zimtohrli} and Multi-Mel SNR~\cite{zang2026msrchallenge}, following the protocol used in the recent Music Source Restoration Challenge.

\subsubsection{Time-domain metric}
We report scale-invariant SNR (SI-SNR) as a time-domain reference metric that is commonly used in enhancement and restoration settings.

\subsubsection{Operational metrics}
We assess efficiency and real-time viability through real-time factor ($\mathrm{RTF}$) and algorithmic latency ($\mathrm{lat}_{\mathrm{alg}}$) with batch size $=1$. We further report model size ($|\theta|$) and measured streaming throughput on CPU and GPU execution paths.

\subsection{Datasets}
\label{subsec:datasets}

\subsubsection{SonicMaster Dataset}
For training and evaluation, we use the SonicMaster dataset~\cite{melechovsky2025sonicmaster}, which contains 168k clean--degraded pairs. It spans ten genre groups (e.g., Hip-Hop) with fine-grained tags. Corruptions are formed by randomly applying one to three degradations (from 19 total) across five categories: EQ, dynamics, reverb, amplitude, and stereo. Each clean track has seven corrupted variants. Following~\cite{melechovsky2025sonicmaster}, we select 1,000 songs and their degradations (7,000 in total) as a test set.

\subsubsection{M\&N Dataset}
We employ the M\&N dataset~\cite{cheng2024musicecan} for evaluation. It contains clean music from nine categories (e.g., piano, harp, song, multi-instrument) mixed with five noise types (electrical, crowd, weather, traffic, stationary). The local paired test split contains 123 clean--noisy examples and is used to evaluate all reported M\&N rows.

\subsubsection{Instrument Datasets}
We also use a training set of solo and ensemble instrument recordings (GuitarSet~\cite{xi2018guitarset}, VocalSet~\cite{wilkins2018vocalset}, SynthSOD~\cite{garcia2024synthsod}, IDMT-PIANO-MM~\cite{abesser2023idmtpianomm}, MAESTRO~\cite{hawthorne2019maestro}, IDMT-SMT-Bass~\cite{abesser2023idmtsmtbass}, FiloBass~\cite{riley2023filobass}) to simulate room recording scenarios. Each clip is degraded online (filtering, amplitude changes, noise, RIR convolution, normalization) using audiomentations~\cite{jordal2024audiomentations}. This dataset is used only in Stage~3.

\subsection{Models and Runtime Protocol}
During training, we use random 2-second audio chunks. Waveforms are resampled to 44.1 kHz and transformed into spectrograms with a window size of 1024 and a hop size of 512. Unless otherwise stated, models are trained with AdamW, a weight decay of $1 \times 10^{-4}$, and a learning rate of $5 \times 10^{-4}$ for each stage. Streaming inference uses a batch size of one, causal state updates, and 1024-sample blocks. The algorithmic latency for all STFT-based models equals the analysis window length, i.e. $\mathrm{lat}_{\mathrm{alg}} = 23.2$~ms at 44.1 kHz. RTF is measured as wall-clock inference time divided by the block duration; $\mathrm{RTF}<1$ therefore indicates real-time processing.

\subsection{Configuration Study}
\label{subsec:config-study}

To validate the curriculum, we conduct two supporting analyses: CRN checkpoints after Stages~1--3 and grouped Stage~3 loss ablations that remove the time-domain, spectral, or music-perceptual block. 

\section{Results}
\label{sec:results}

Table~\ref{tab:runtime} addresses the real-time requirement directly. With a batch size of one and causal 1024-sample blocks, all four causal models process a 23.2~ms block in 2.46--2.65~ms on the tested GPU, giving $\mathrm{RTF}=0.106$--$0.114$. Thus, the evaluated systems are both frame-causal and faster than real time on the tested accelerator. CPU diagnostics are less uniform, especially for CRN, indicating that reliable CPU-only deployment would require optimized kernels.

Table~\ref{tab:mnm_sonicmaster_metrics} shows that no single method dominates all metrics or datasets. Among the displayed Stage~3 speech-derived causal systems, CRN gives the strongest MM-SNR and SI-SNR on both datasets; DFN-MEL is the next strongest on these two M\&N measures but transfers poorly to the broader SonicMaster degradations. MFN-MS is not the overall best model: on M\&N it matches the degraded-input FAD and improves ZIM, whereas on SonicMaster it slightly improves KL and SSIM but worsens MM-SNR, SI-SNR, PQ, and ZIM. The offline references are similarly task-dependent. MusicECAN is strongest on several M\&N metrics, consistent with its denoising target, whereas SonicMaster is strongest on SonicMaster FAD and KL but not on the signal-preservation metrics.

\begin{table}[ht] 
    \centering
    \caption{Streaming runtime evidence at 44.1 kHz with a batch size of one and 1024-sample causal blocks. CPU and GPU columns report mean inference time in ms per 23.2~ms block. CPU values are unoptimized local PyTorch diagnostics; the real-time claim is based on the measured streaming GPU path, where all models have $\mathrm{RTF}<1$.}    
    \scriptsize
    \setlength{\tabcolsep}{2.2pt}
    \renewcommand{\arraystretch}{1.05}
    \begin{tabular}{lrrrrr}
    \toprule
    \textbf{Model} & \textbf{M} & \textbf{CPU} & \textbf{GPU} & \textbf{RTF} & \textbf{Lat.} \\
    \midrule 
      DFN &  1.5 & 8.74 & 2.65 & 0.114 & 23.2 \\
      DFN-MEL &  1.5 & 8.74 & 2.65 & 0.114 & 23.2 \\
      CRN & 5.6 & 69.20 & 2.56 & 0.110 & 23.2 \\
      MFN-MS & 16.5 & 17.56 & 2.46 & 0.106 & 23.2 \\
    \bottomrule
    \end{tabular}
    \label{tab:runtime}
\end{table}

The Supplementary Material reports the omitted checkpoints and diagnostics behind this interpretation. The stage-wise rows show that CRN Stage~2 improves SonicMaster FAD/KL relative to Stage~1, while Stage~3 shifts the model toward MM-SNR/SI-SNR. The grouped loss ablation identifies the time-domain block as the main driver of M\&N SI-SNR/MM-SNR gains, and the MFN-MS category breakdown shows that its largest negative MM-SNR delta occurs for stereo-image degradations. These results reinforce that metric families should be compared rather than collapsed into a single ranking.

The observed metric disagreement is expected and informative rather than a defect of the benchmark: waveform, spectrogram, embedding, production-quality, and psychoacoustic measures encode different notions of fidelity and different invariances~\cite{roblek2019fad,jepsen2025sisdr_noisyrefs,gui2023fad_adapt,kad2025}. We therefore regard agreement across metric families as stronger evidence than an isolated gain, while disagreement identifies cases in which nuisance removal may trade against preservation of the produced mix. Without subjective data, we do not claim that any one metric predicts listener preference.

\section{Conclusion}
\label{sec:conclusion}

We presented a benchmark and diagnostic analysis of causal neural models for real-time music enhancement. Runtime measurements show that compact STFT-based systems can operate faster than real time at 44.1~kHz on the tested GPU. However, no evaluated model consistently improves the degraded input across both datasets and all metric families. The central result is therefore not a claim of MFN-MS superiority: causal music enhancement is computationally feasible, but indiscriminate always-on processing can worsen the input under several objective criteria and may trade nuisance removal against preservation of the intended production. The category-level diagnostics, including the negative stereo MM-SNR result, motivate degradation-aware routing, stereo-aware processing, and an identity-preserving do-no-harm path that defaults to passthrough when correction is uncertain. Subjective evaluation remains important future work for determining which objective trade-offs correspond to listener preference.

\clearpage

\section*{Aknowledgments}
This work has been supported by the European Research Council (ERC) under the EU's Horizon 2020 research \& innovation programme, grant agreement No. 101019375 (Whither Music?).

\bibliographystyle{IEEEtran}
\bibliography{biblio}

@inproceedings{crn,
  title={A convolutional recurrent neural network for real-time speech enhancement.},
  author={Tan, Ke and Wang, DeLiang},
  booktitle={Interspeech},
  volume={2018},
  pages={3229--3233},
  year={2018}
}

@inproceedings{hu2020dccrn,
  title={DCCRN: Deep complex convolution recurrent network for phase-aware speech enhancement},
  author={Hu, Yanxin and Liu, Yun and Lv, Shubo and Xing, Mengtao and Zhang, Shimin and Fu, Yihui and Wu, Jian and Zhang, Bihong and Xie, Lei},
  booktitle={Interspeech},
  year={2020}
}

@inproceedings{schroter2023deepfilternet,
  title={Deepfilternet: Perceptually motivated real-time speech enhancement},
  author={Schr{\"o}ter, Hendrik and Rosenkranz, Tobias and Escalante-B, Alberto N and Maier, Andreas},
  booktitle={Interspeech},
  year={2023}
}

@inproceedings{babaev2024finally,
  title     = {FINALLY: fast and universal speech enhancement with studio-like quality},
  author    = {Babaev, Nicholas and Tamogashev, Kirill and Saginbaev, Azat and Shchekotov, Ivan and Bae, Hanbin and Sung, Hosang and Lee, WonJun and Cho, Hoon-Young and Andreev, Pavel},
  booktitle = {Advances in Neural Information Processing Systems (NeurIPS)},
  year      = {2024}
}

@article{melechovsky2025sonicmaster,
  title={SonicMaster: Towards Controllable All-in-One Music Restoration and Mastering},
  author={Melechovsky, Jan and Mehrish, Ambuj and Herremans, Dorien},
  journal={arXiv preprint arXiv:2508.03448},
  year={2025}
}

@article{moliner2022behm,
  title={BEHM-GAN: Bandwidth extension of historical music using generative adversarial networks},
  author={Moliner, Eloi and V{\"a}lim{\"a}ki, Vesa},
  journal={IEEE/ACM Transactions on Audio, Speech, and Language Processing},
  volume={31},
  pages={943--956},
  year={2022},
  publisher={IEEE}
}

@inproceedings{moliner2022two,
  title={A two-stage u-net for high-fidelity denoising of historical recordings},
  author={Moliner, Eloi and V{\"a}lim{\"a}ki, Vesa},
  booktitle={International Conference on Acoustics, Speech and Signal Processing (ICASSP)},
  pages={841--845},
  year={2022},
  organization={IEEE}
}

@article{cheng2024musicecan,
  title={MusicECAN: An Automatic Denoising Network for Music Recordings With Efficient Channel Attention},
  author={Cheng, Haonan and Liu, Shulin and Lian, Zhicheng and Ye, Long and Zhang, Qin},
  journal={IEEE/ACM Transactions on Audio, Speech, and Language Processing},
  volume={32},  
  year={2024},  
}

@inproceedings{xi2018guitarset,
  title={GuitarSet: A Dataset for Guitar Transcription.},
  author={Xi, Qingyang and Bittner, Rachel M and Pauwels, Johan and Ye, Xuzhou and Bello, Juan Pablo},
  booktitle={Proceedings of the International Conference of Music Information Retrieval (ISMIR)},
  pages={453--460},
  year={2018}
}

@inproceedings{roblek2019fad,
  title     = {Fr{\'{e}}chet Audio Distance: A Reference-Free Metric for Evaluating Music Enhancement Algorithms},
  author    = {Kilgour, Kevin and Zuluaga, Mauricio and Roblek, Dominik and Sharifi, Matthew},
  booktitle = {Proc. Interspeech},
  pages     = {2350--2354},
  year      = {2019},
}

@article{gui2023fad_adapt,
  title   = {Adapting Fr{\'{e}}chet Audio Distance for Generative Music Evaluation},
  author  = {Gui, Azalea and Gamper, Hannes and Braun, Sebastian and Emmanouilidou, Dimitra},
  journal = {IEEE International Conference on Acoustics, Speech, and Signal Processing (ICASSP)},
  year    = {2024}
}

@inproceedings{jepsen2025sisdr_noisyrefs,
  title   = {A Study of the Scale-Invariant Signal-to-Distortion Ratio in Speech Separation with Noisy References},
  author  = {Jepsen, Simon Dahl and Christensen, Mads Gr{\ae}sb{\o}ll and Jensen, Jesper Rindom},
  booktitle = {IEEE ASRU, Workshop on Automatic Speech Recognition and Understanding.},
  year    = {2025},  
}

@inproceedings{kad2025,
  title   = {KAD: No More FAD! An Effective and Efficient Evaluation Metric for Audio Generation},
  author  = {Chung, Yoonjin and Eu, Pilsun and Lee, Junwon and Choi, Keunwoo and Nam, Juhan and Chon, Ben Sangbae},
  booktitle = {AI Heard That! ICML Workshop on Machine Learning for Audio},
  year    = {2025},  
}

@dataset{wilkins2018vocalset,
  author       = {Wilkins, Julia and Seetharaman, Prem and Wahl, Alison and Pardo, Bryan},
  title        = {VocalSet: A Singing Voice Dataset},
  publisher    = {Zenodo},
  year         = {2018},
  doi          = {10.5281/zenodo.1442513}
}

@inproceedings{hawthorne2019maestro,
  title     = {Enabling Factorized Piano Music Modeling and Generation with the MAESTRO Dataset},
  author    = {Hawthorne, Curtis and Stasyuk, Andriy and Roberts, Adam and Simon, Ian and Huang, Cheng-Zhi Anna and Dieleman, Sander and Elsen, Erich and Engel, Jesse and Eck, Douglas},
  booktitle = {International Conference on Learning Representations (ICLR)},
  year      = {2019}
}

@dataset{abesser2023idmtpianomm,
  author       = {Abe{\ss}er, Jakob and Bittner, Franca and Richter, Maike and Gonzalez, Marcel and Lukashevich, Hanna},
  title        = {IDMT-PIANO-MM Dataset},
  publisher    = {Zenodo},
  year         = {2023},
  doi          = {10.5281/zenodo.7544248}
}

@dataset{abesser2023idmtsmtbass,
  author       = {Abe{\ss}er, Jakob},
  title        = {IDMT-SMT-Bass Dataset},
  publisher    = {Zenodo},
  year         = {2023},
  doi          = {10.5281/zenodo.7188892}
}

@article{garcia2024synthsod,
  title   = {SynthSOD: Developing an Heterogeneous Dataset for Orchestra Music Source Separation},
  author  = {Garc{\'{\i}}a-Mart{\'{\i}}nez, Jaime and D{\'{\i}}az-Guerra, David and Politis, Archontis and Virtanen, Tuomas and Carabias-Orti, Julio J. and Vera-Candeas, Pedro},
  journal = {IEEE Open Journal of Signal Processing},
  year    = {2024}
}

@inproceedings{riley2023filobass,
  title   = {FiloBass: A Dataset and Corpus Based Study of Jazz Basslines},
  author  = {Riley, Xavier and Dixon, Simon},
  booktitle = {International Symposium of Music Information Retrieval (ISMIR)},
  year    = {2023},
  note    = {Dataset available via Zenodo: 10.5281/zenodo.10069709}
}

@article{jordal2024audiomentations,
  title={Audiomentations documentation},
  author={Jordal, I},
  journal={Github. io},
  year={2024}
}

@article{tjandra2025aes,
    title={Meta Audiobox Aesthetics: Unified Automatic Quality Assessment for Speech, Music, and Sound},
    author={Andros Tjandra and Yi-Chiao Wu and Baishan Guo and John Hoffman and Brian Ellis and Apoorv Vyas and Bowen Shi and Sanyuan Chen and Matt Le and Nick Zacharov and Carleigh Wood and Ann Lee and Wei-Ning Hsu},
    year={2025},
    url={https://arxiv.org/abs/2502.05139}
}

@inproceedings{elizalde2023clap,
  title={Clap learning audio concepts from natural language supervision},
  author={Elizalde, Benjamin and Deshmukh, Soham and Al Ismail, Mahmoud and Wang, Huaming},
  booktitle={International Conference on Acoustics, Speech and Signal Processing (ICASSP)},
  pages={1--5},
  year={2023},
  organization={IEEE}
}

@article{drgas2023survey,
  title={A survey on low-latency DNN-based speech enhancement},
  author={Drgas, Szymon},
  journal={Sensors},
  volume={23},
  number={3},
  pages={1380},
  year={2023},
  publisher={MDPI}
}

@article{alakuijala2025zimtohrli,
  title   = {Zimtohrli: An Efficient Psychoacoustic Audio Similarity Metric},
  author  = {Alakuijala, Jyrki and Bruse, Martin and Boukortt, Sami and Marus Coldenhoff, Jozef and Cernak, Milos},
  journal = {arXiv preprint arXiv:2509.26133},
  year    = {2025},
  doi     = {10.48550/arXiv.2509.26133}
}

@article{zang2026msrchallenge,
  title   = {Summary of The Inaugural Music Source Restoration Challenge},
  author  = {Zang, Yongyi and Hai, Jiarui and Ge, Wanying and Kong, Qiuqiang and Dai, Zheqi and Wang, Helin and Mitsufuji, Yuki and Plumbley, Mark D.},
  journal = {arXiv preprint arXiv:2601.04343},
  year    = {2026},
  doi     = {10.48550/arXiv.2601.04343},
  note    = {ICASSP 2026 Music Source Restoration Challenge protocol and metrics (Multi-Mel-SNR, Zimtohrli, FAD-CLAP).}
}

\singlepagebreak
\onecolumn
\appendix[Supplementary Material]
\setcounter{table}{0}
\renewcommand{\thetable}{S\arabic{table}}

This appendix provides the complete results behind the compact comparison in the main letter. The main paper keeps the degraded input, final Stage~3 checkpoints, MFN-MS, and the two offline references to emphasize the deployable setting. Here, we include the intermediate checkpoints and diagnostics used to interpret those rows. The metrics are the same as in the main paper: lower is better for FAD/KL, and higher is better for MM-SNR, SI-SNR, SSIM, PQ, and Zimtohrli (ZIM).

\begin{table}[ht]
\caption{Complete evaluation on M\&N and SonicMaster datasets.}
\label{tab:supp_complete}
\centering
{\setlength{\tabcolsep}{2pt}
\begin{tabular}{lrrrrrrr|rrrrrrr}
\toprule
& \multicolumn{7}{c|}{\textbf{M\&N Dataset}} & \multicolumn{7}{c}{\textbf{SonicMaster Dataset}} \\
\cmidrule(lr){2-8}\cmidrule(lr){9-15}
\textbf{Model} & FAD$\downarrow$ & KL$\downarrow$ & MM$\uparrow$ & SI-SNR$\uparrow$ & SSIM$\uparrow$ & PQ$\uparrow$ & ZIM$\uparrow$ & FAD$\downarrow$ & KL$\downarrow$ & MM$\uparrow$ & SI-SNR$\uparrow$ & SSIM$\uparrow$ & PQ$\uparrow$ & ZIM$\uparrow$ \\
\midrule
Degraded input & 0.640 & 0.152 & 5.336 & 3.509 & 0.537 & 5.818 & 2.391 & 0.103 & 0.077 & 8.477 & 0.023 & 0.501 & 7.007 & 3.659 \\
\midrule
CRN Stage~1 & 0.663 & 0.077 & 2.685 & -0.515 & 0.475 & 5.983 & 2.307 & 0.089 & 0.325 & 4.420 & -2.185 & 0.555 & 7.056 & 3.725 \\
CRN Stage~2 & 0.688 & 0.069 & 2.848 & -0.650 & 0.481 & 6.005 & 2.287 & 0.085 & 0.186 & 4.021 & -2.474 & 0.558 & 7.064 & 3.726 \\
CRN Stage~3 & 0.684 & 0.099 & 7.048 & 4.556 & 0.502 & 5.903 & 2.339 & 0.102 & 0.350 & 5.511 & -1.990 & 0.508 & 6.988 & 3.657 \\
\midrule
DFN Stage~1 & 0.702 & 0.162 & 1.283 & -2.830 & 0.427 & 5.848 & 2.285 & 0.115 & 0.199 & 3.383 & -5.112 & 0.537 & 6.846 & 3.507 \\
DFN Stage~2 & 0.724 & 0.144 & 1.225 & -3.266 & 0.421 & 5.882 & 2.283 & 0.117 & 0.189 & 3.385 & -4.962 & 0.544 & 6.865 & 3.510 \\
DFN Stage~3 & 0.655 & 0.184 & 4.001 & 1.929 & 0.465 & 5.742 & 2.338 & 0.114 & 0.205 & 3.296 & -4.410 & 0.503 & 6.869 & 3.528 \\
\midrule
DFN-MEL Stage~1 & 0.727 & 0.168 & 2.503 & 0.358 & 0.444 & 5.892 & 2.267 & 0.144 & 0.228 & 2.886 & -5.025 & 0.494 & 6.582 & 3.306 \\
DFN-MEL Stage~2 & 0.751 & 0.128 & 5.351 & 3.333 & 0.513 & 5.476 & 2.208 & 0.120 & 0.285 & 3.587 & -4.228 & 0.575 & 6.915 & 3.556 \\
DFN-MEL Stage~3 & 0.754 & 0.132 & 5.348 & 3.898 & 0.512 & 5.475 & 2.223 & 0.145 & 0.222 & 2.794 & -5.179 & 0.494 & 6.564 & 3.306 \\
\midrule
MFN-MS Stage~1 & 0.728 & 0.108 & -0.422 & -12.887 & 0.459 & 6.083 & 2.223 & 0.232 & 0.035 & -1.585 & -14.995 & 0.557 & 7.089 & 2.231 \\
MFN-MS Stage~2 & 0.671 & 0.177 & 3.658 & 1.636 & 0.493 & 5.675 & 2.441 & 0.149 & 0.071 & 3.323 & -6.353 & 0.507 & 6.876 & 2.872 \\
MFN-MS Stage~3 & 0.640 & 0.190 & 4.284 & 2.689 & 0.469 & 5.657 & 2.428 & 0.116 & 0.074 & 4.330 & -2.877 & 0.503 & 6.878 & 3.271 \\
\midrule
MusicECAN (offline) & 0.644 & 0.056 & 8.596 & 5.532 & 0.573 & 5.904 & 2.376 & 0.188 & 0.082 & 4.457 & -3.073 & 0.447 & 6.843 & 3.429 \\
SonicMaster (offline) & 0.697 & 0.124 & 2.654 & -1.150 & 0.456 & 5.898 & 2.296 & 0.072 & 0.008 & 2.021 & -3.309 & 0.519 & 3.028 & 2.885 \\
\bottomrule
\end{tabular}}
\end{table}

Table~\ref{tab:supp_complete} expands the compact main-table view by adding the Stage~1 and Stage~2 checkpoints for both evaluation sets. It should be read as a trace of changes across training rather than as a single leaderboard. CRN Stage~3 gives the strongest speech-derived real-time result on MM-SNR/SI-SNR, while CRN Stage~2 is better on SonicMaster FAD/KL and PQ/ZIM. DFN-MEL remains competitive on M\&N signal-preservation metrics but weakens on the broader SonicMaster degradations, which suggests that a mel-domain representation alone is not sufficient for general music restoration. The offline references are also task-dependent: MusicECAN is strongest on the denoising-oriented M\&N split, whereas SonicMaster is strongest only on SonicMaster FAD/KL among the displayed metric families.

Table~\ref{tab:supp_diagnostics} collects the two diagnostics used to interpret these trends. Panel (a) removes one Stage~3 loss block at a time from the same CRN Stage~2 checkpoint. The time-domain block contains SI-SDR and PE-SI-SDR, the spectral block contains MR-STFT terms, and the music-perceptual block contains log-mel and instantaneous-frequency regularization. Panel (b) reports MFN-MS deltas over the degraded input on SonicMaster; positive values indicate improvement, and counts can exceed the number of examples because one clip may contain multiple degradation categories. Together, the diagnostics show that time-domain losses drive the largest SI-SNR/MM-SNR gains, while the near-neutral non-stereo rows and the negative stereo MM-SNR delta point to the need for stereo-aware and degradation-aware correction.

\begin{table}[ht]
\caption{Diagnostic analyses: (a) CRN Stage~3 grouped loss ablations on M\&N; (b) MFN-MS SonicMaster deltas by degradation category.}
\label{tab:supp_diagnostics}
\centering
\begin{tabular}{cc}
\begin{tabular}{lrrrr}
\toprule
\multicolumn{5}{c}{\textbf{(a) Loss ablation}} \\
\midrule
\textbf{Metric} & \textbf{Full} & \textbf{--Music} & \textbf{--Spec.} & \textbf{--Time} \\
\midrule
FAD$\downarrow$ & 0.684 & \textbf{0.670} & 0.705 & 0.687 \\
KL$\downarrow$ & \textbf{0.099} & 0.174 & 0.197 & 0.105 \\
MM$\uparrow$ & \textbf{7.048} & 6.035 & 6.580 & 3.952 \\
SI$\uparrow$ & \textbf{4.556} & 3.976 & 4.312 & 1.783 \\
SSIM$\uparrow$ & 0.502 & 0.503 & 0.480 & \textbf{0.537} \\
PQ$\uparrow$ & \textbf{5.904} & 5.855 & 5.776 & 5.826 \\
ZIM$\uparrow$ & 2.339 & 2.335 & 2.317 & \textbf{2.353} \\
\bottomrule
\end{tabular}
&
\begin{tabular}{lrrrr}
\toprule
\multicolumn{5}{c}{\textbf{(b) MFN-MS category deltas}} \\
\midrule
\textbf{Category} & \textbf{N} & SI$_i$ & SNR$_i$ & MM$_i$ \\
\midrule
Amplitude & 1565 & 0.0187 & 0.0960 & 0.0224 \\
Dynamics & 1678 & 0.0006 & 0.0096 & 0.0005 \\
EQ & 4062 & 0.0006 & 0.0130 & 0.0011 \\
Reverb & 2696 & -0.0002 & 0.0116 & 0.0000 \\
Stereo & 999 & 0.0041 & 0.0211 & -0.0725 \\
\bottomrule
\end{tabular}
\end{tabular}
\end{table}

\end{document}